 \documentclass[prd,twocolumn,groupedaddress,showpacs]{revtex4}
\usepackage[dvips]{graphicx}
\usepackage[]{caption}
\usepackage{amsmath}
\usepackage{amssymb}
\usepackage{dcolumn}
\usepackage{latexsym}
\newcommand{\beq}{\begin{equation}}
\newcommand{\eeq}{\end{equation}}
\newcommand{\be}{\begin{eqnarray}} 
\newcommand{\en}{\end{eqnarray}}
\newcommand{\ep}{\epsilon}

\newcommand{\CA}{\hbox{{$\cal A$}}}

\newcommand{\CH}{\hbox{{$\cal H$}}}
\newcommand{\CJ}{\hbox{{$\cal J$}}}

\newcommand{\et}{|\eta|}
\newcommand{\eti}{|\eta_{i}|}
\newcommand{\etu}{|\eta_{1}|}
\newcommand{\etd}{|\eta_{2}|}

\begin{document}
\bibliographystyle{prsty}
\title{T-Duality and the Spectrum of Gravitational Waves}
\author{Paul Robert Chouha $^{1)}$\footnote{paulchouha@hep.physics.mcgill.ca}
and Robert H. Brandenberger $^{1,2)}$\footnote{rhb@hep.physics.mcgill.ca}}
\affiliation{1) Physics Department, McGill University, 3600 rue Universit\'e, 
Montreal, QC H3A 2T8, CANADA}
\affiliation{2) Perimeter Institute for Theoretical Physics, 31 Caroline St. N.,
Waterloo, ON, N2L 2Y5, CANADA}
\date{\today}
\preprint{hep-th/0508119}
\begin{abstract}

In the inflationary universe scenario, the physical wavelength of 
cosmological fluctuation modes which are currently probed
in observations was shorter than the Hubble radius, and in fact
shorter than the Planck and string lengths, at the beginning of the
period of inflation. Thus, during the early stages of evolution,
the fluctuations are subject to Planck scale physics. 
In the context of an inflationary cosmological background, we
examine the signatures of a specific modified dispersion relation  
motivated by the T-duality symmetry of string
theory on the power spectrum of gravitational waves. 
The modified dispersion relation is extracted from the 
asymptotic limit of the string center of mass propagator.

\end{abstract}
\maketitle

\section{Introduction}

The inflationary universe acts as a microscope which allows us, by
means of cosmological observations performed today, to probe scales
which at the beginning of the period of inflation were sub-Hubble
(i.e. the wavelength was smaller than the Hubble radius), and in
fact sub-Planckian (wavelength smaller than the Planck scale)
\cite{RHBrev}. This is possible since during the period of inflation
(which we for concreteness take to be almost exponential) the physical
wavelength of fixed perturbation modes increases exponentially, whereas
the Hubble radius remains approximately constant. Thus, in principle,
it is possible to probe Planck-scale or string-scale physics using
current observations. 

Crucial to the study of trans-Planckian effects on current cosmological
observations is the ability to track the cosmological fluctuations from
the earliest times and smallest scales to late times. Signatures of
trans-Planckian physics then emerge via encoding the new physics which
is effective at the Planck scale on the initial conditions and early
evolution of the linearized fluctuations. This program was initially carried out
\cite{Jerome1,Niemeyer} making use of ad-hoc modified dispersion relations
for the cosmological perturbations. These dispersion relations are linear
(i.e. the usual ones) for physical wavenumbers smaller than a cutoff
value which corresponds to the energy scale of the new physics, and deviates
for larger wavenumbers. This approach has the advantage
that it is easily computationally tractable. It was shown \cite{Jerome1} that
observable effects are possible provided that the dispersion relation
violates the adiabaticity condition, i.e. that the early evolution of the
fluctuations is non-adiabatic from the point of view of the evolution which
would have been obtained had the dispersion relation been the un-modified one.
There are some constraints on the magnitude of trans-Planckian effects
which come from back-reaction analyses \cite{Jerome2}, but these are
less severe than initially conjectured in \cite{Tanaka,Starob}.

There are other approaches to studying the possible imprints of trans-Planckian
physics on cosmological observations. In a ``minimal'' approach, initial
conditions on the fluctuations are imposed not at some early initial time which
is the same for all wavenumbers, but for each wavenumber at the time when the
corresponding wavelength equals the scale of the new physics \cite{minimal}.
There are approaches which attempt to include crucial features of candidate
theories of quantum gravity, e.g. via space-space uncertainty relations
\cite{Columbia}, space-time uncertainty relations \cite{Ho}, Q-deformed
structures \cite{Sera}, and minimal length effects \cite{Kempf}. Analyses
based on an effective field theory approach have also been performed
\cite{Cliff} (see \cite{Jerome4} for a review and a more complete list
of references).  
 
In this paper we work under the hypothesis that superstring theory is the
correct description of physics on the smallest length scales. A key
feature which distinguishes string theory from point particle-based quantum
field theory is the new symmetry of T-duality \cite{Tduality}. We will
explain this symmetry in the context of a theory with closed strings only,
although the symmetry extends to theories with open strings \cite{Pol,Boehm}.
Let us assume that each spatial direction is a torus with radius $R$. Closed
strings have various degrees of freedom: momentum modes which correspond
to the center of mass motion of the strings and whose energies are quantized
in units of $1/R$ (setting the string length equal to 1 for this discussion),
oscillatory modes which correspond to wiggles of the strings and whose energies
are independent of $R$, and winding modes which count the number of times the
string winds the torus, whose energies are quantized in units of $R$.

The string mass spectrum is
\beq
\label{E:1}
M^{2} \, = \, \frac{n^{2}}{R^{2}} + w^{2}R^{2} +
2\left(\tilde{N^{\bot}} + N^{\bot} - 2\right)
\eeq
where $n$ is the integer which determines the Kaluza-Klein momentum modes, 
$w$ is the string winding number, and $N^{\bot}$ and $\tilde{N^{\bot}}$ are 
integers which denote the number of excited oscillation modes on a closed string 
in the right-moving and left-moving directions around the string. Note
that in the case of several compact dimensions, $n$ and $w$ are vectors
of integers, and the squares $n^2$ and $w^2$ appearing above are scalar
products of the respective vectors. The quantum
numbers are constrained by the level matching condition
$\tilde{N^{\bot}}-N^{\bot}=nw$. In what follows we ignore the last term which is
 only a constant when only  the leading terms with $n,w=1$ are considered.
 
T-duality is the transformation which takes the compactified radius to its 
inverse, and exchanges the quantum numbers $n$ and $w$:
\beq 
 R \longleftrightarrow  {1 \over R} \, , \,\,\,\,\, n \longleftrightarrow w \, .
\eeq
The mass spectrum is invariant under this transformation. The T-duality
symmetry has been used \cite{BV} to argue that a stringy early universe
will be nonsingular and the special effects of winding modes could also
explain why there are only three large spatial dimensions (see also
\cite{Perlt} for an early paper, and \cite{ABE,BEK} for more
recent developments). This approach to the stringy early universe is
called ``string gas cosmology''.  

To date, effects of T-duality on the spectrum
of gravitational waves have not been studied. It is the aim
of this paper to perform this study in the context of the hypothesis
that there was a period of cosmological inflation. Note that at the
moment one of the outstanding challenges in the field of string gas
cosmology is to obtain a period of inflation (see \cite{Moshe,Damien,Tirtho}
for some recent ideas).

In what follows we propose to study the effects of T-duality on the power 
spectrum of a test scalar field on an expanding background (the
equation of such a test scalar field is equivalent to the equation
satisfied by the polarization modes of gravitational waves).
We will begin with a modified
propagator which encodes the consequences of T-duality \cite{Smailagic:2003hm}
and Fourier transform to obtain a modified dispersion relation. We then
use the techniques of \cite{Jerome1,Jerome3} to follow the
perturbations from the initial time (when we assume that they start out
in the state that minimizes the Hamiltonian) until they cross the
Hubble radius. From then on, their evolution is standard (see e.g.
\cite{MFB} for a review of the theory of cosmological perturbations and
\cite{RHBrev2} for a pedagogical overview).
  
\section{Preliminaries: A New Modified Dispersion Relation}

Our starting point is the result of \cite{Smailagic:2003hm}, which shows that 
the T-duality symmetry results in the following propagator of the string center 
of mass (after Fourier transforming): 
\beq
 \label{E:2}
 G_{reg}\left(\,k^2\,\right) \,  \approx \,  \frac{1}{(2\pi)^2} 
\frac{l_s K_1\left(\, l_s\sqrt{\,k^2+m^2}\,\right) }{\,\sqrt{\,k^2 + m^2}}
\eeq
where $K_{n}(z)$ is the modified Bessel function of order $n$, and $l_s$ is
the length scale below which trans-Planckian physics is important (in our
case it is the string length). This 
is the same result found earlier in  \cite{padma2} under the assumption that 
quantum gravity effects lead to a modification of the space-time interval on
short distance scales, the modified formula for the interval being
$ (x-y)^{2}+l_{s}^{2}$ instead of $(x-y)^{2}$ .  

Making use of the asymptotic forms of the Bessel functions, the propagator 
has the following limiting forms
\begin{eqnarray}
\label{E:3}
G_{reg}\left(k^2\right)\rightarrow \ 
\begin{cases} 
\displaystyle{\frac{1}{k^2+m^2}}\, ,& 
\mbox{$l_s\sqrt{k^2+m2}\ll 1$}\, , 
\cr 
 & \cr 
\exp (-l_s\sqrt{k^2+m^2})\, , & \mbox{$l_s\sqrt{k^2+m2}\gg 1$} \, . 
\end{cases}
\end{eqnarray}
As is evident, for momenta small compared to $l_s^{-1}$, the propagator approaches
the usual one, whereas for large wavenumbers it decays exponentially with $k$. This
decay is a reflection of the fact that T-duality smoothes out the usual ultraviolet
divergences of the theory. Writing the propagator in momentum space as $\omega^{-2}$,
where $\omega$ is the effective frequency, we can read off from (\ref{E:3}) the
modified dispersion relation. In the high energy regime $l_s\sqrt{k^2+m2}\gg 1$
it takes the form
\beq
\label{E:4}
\omega \, =  \, F(k) \, = \, \frac{1}{l_s}\exp (l_s\sqrt{k^2+m^2}) \, .
\eeq
Note that in the above (\ref{E:2}), (\ref{E:3}) and (\ref{E:4}), $k$ and
$\omega$ denote the physical momenta and frequencies and not the comoving 
quantities. The comoving wavenumber will be denoted by $n$.

In cosmology we follow the evolution of fluctuations which correspond to
waves with a fixed comoving wavenumber. For these waves, the
dispersion relation is time dependent. Instead of the simple linear form 
\beq
\omega^2 \, = \,  k^2 \, \equiv \, \frac{n^2}{a^2}
\eeq
it takes the form
\beq \label{freq}
\omega \, = \, F\left[\frac{n}{a(\eta)}\right] \, ,
\eeq
where $\eta$ denotes conformal time, in terms of which the background metric is
\beq
ds^2 \, = \, a^2(\eta) \bigl[ d\eta^2 - d{\bf x}^2 \bigr] \, ,
\eeq
$d{\bf x}^2$ denoting the metric of flat Euclidean three space, and $a(\eta)$
being the scale factor of the universe.
 
To study the effects this modified dispersion relation has on the power spectrum of 
gravitational waves, we follow the approach of \cite{Jerome1}.
We start with a free massless scalar field $\phi(\eta,\bf{x})$ living in the
above background space-time. The Fourier modes
evolve independently, as do the Fourier modes of the cosmological
perturbations in linear theory. After introducing the 
re-scaled field $\mu$ defined via 
\beq
\phi(\eta,\bf{x}) \, \equiv \, \frac{1}{(2\pi)^{3/2}}\int d \bf{n}
(\mu/a)e^{i\bf{n}.\bf{x}} \, ,
\eeq
the equation of motion for the Fourier modes of $\mu$ reads
\beq \label{eom1}
\mu^{\prime\prime}+\left[n^2-\frac{a^{ \prime\prime}}{a}\right]\mu \, = \, 0 \, .
\eeq
Note that gravitational waves in an expanding space obey precisely this
equation, whereas the scalar metric fluctuations obey an equation which is
very similar. The only change is that the time-dependent negative square
mass $-\frac{a^{ \prime\prime}}{a}$ in the above equation is replaced by
$-\frac{z^{ \prime\prime}}{z}$, where $z(\eta)$ is a more general function
of the cosmological background (which, in fact, is proportional to $a(\eta)$ if
the equation of state of the background does not change in time -
see e.g. \cite{MFB,RHBrev2} for reviews).

To take into account the modification of the dispersion relation,
we need to replace (\ref{eom1}) by
\beq
\label{EOM}
\mu^{\prime\prime}+\left[n_{eff}^2-\frac{a^{ \prime\prime}}{a}\right]\mu \, = \, 0
\, ,
\eeq
where the effective comoving frequency $n_{eff}$ is given by
\beq
\label{E:8}
n_{eff}^2 \, = \, a^2(\eta)F^2[n/a(\eta)]
\eeq
\begin{widetext}
\subsection{Trans-Planckian Region}\label{S:TransPlkI}
In an inflationary background, any Fourier mode passes through three 
different intervals of time which are depicted in Figure 1. These
regions are:
\begin{figure*}
\includegraphics[width=.99\textwidth,height=.90\textwidth]{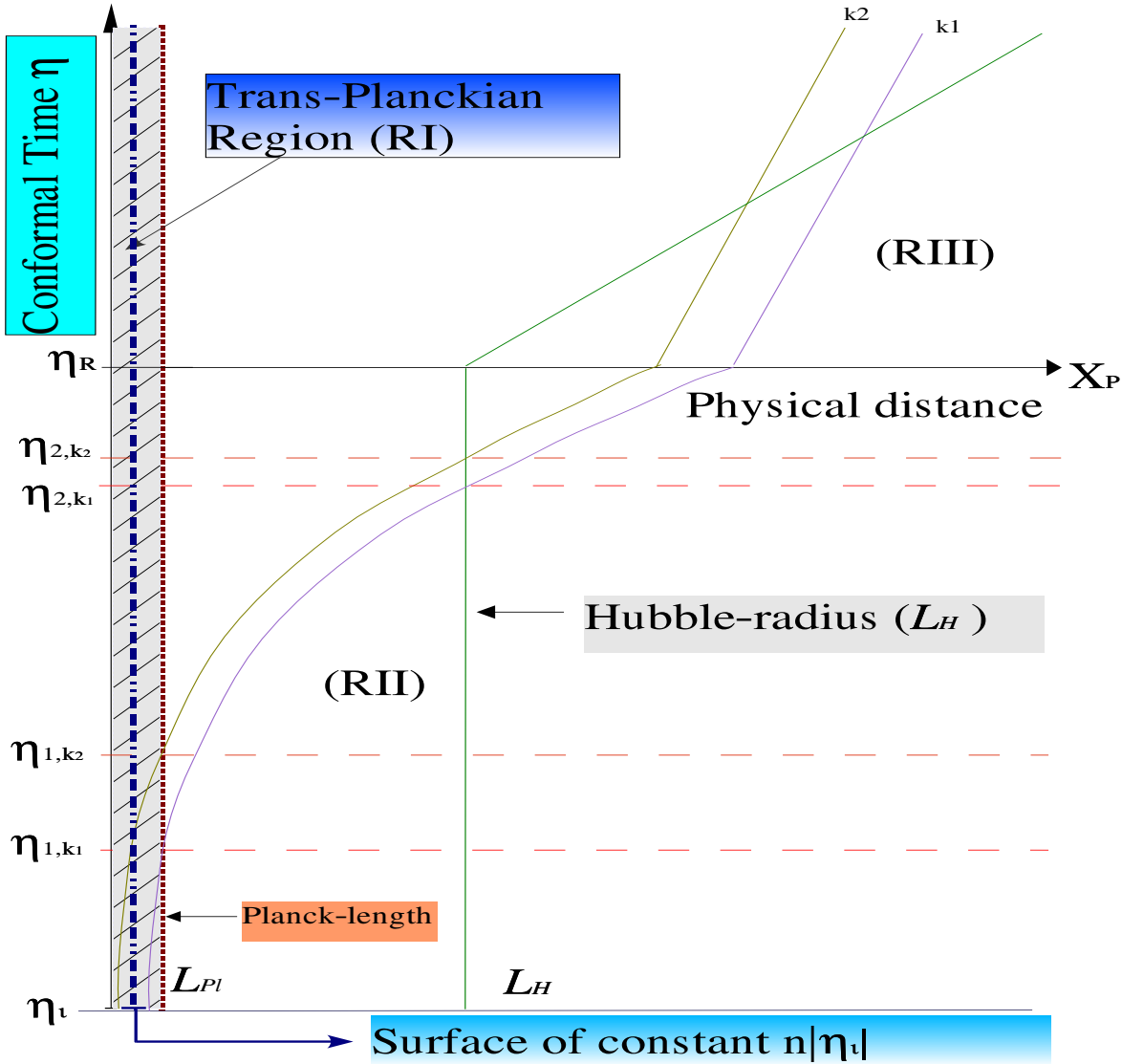}
\caption{Sketch of the relevant regions for our model including the 
Trans-Planckian region where the WKB approximation for the mode
equation breaks down. The horizontal axis is physical distance, the
vertical axis conformal time. The red (dotted) line indicates the
Planck or string length, the distance where the usual dispersion relation
becomes invalid. The green (solid) line which is vertical during the
period of inflation which lasts from $\eta_i$ to $\eta_R$, is the
boundary between Regions II and III, the regions where the mode evolution
is oscillatory or frozen, respectively. The two thin lines for which the
distance grows exponentially during the period of inflation indicate the
physical wavelengths of two fixed comoving modes $k_1$ and $k_2$. As
is evident, they start out in Region I, the Trans-Planckian region, at early
times, and then cross Region II to end up in Region III during the later
stages of inflation.}  
\label{disp}
\end{figure*}
\end{widetext}
\begin{itemize}
\item[\textbf{(RI)}] \textbf{Region I}: when the physical wavelength of a given mode, 
$\lambda(\eta) \equiv (2\pi/n)a(\eta)$ is smaller than the cutoff length where the
dispersion relation begins to deviate from its usual linear form. Here, it is
taken to be the string length $l_s$: $\lambda < l_{S}$.
\item[\textbf{(RII)}] \textbf{Region II}: when the wavelength is longer than $l_s$
but smaller than the Hubble radius: $l_{S}<\lambda<l_{H}$
\item[\textbf{(RIII)}] \textbf{Region III}: when the wavelength exceeds $l_{H}$, the
mode has frozen out and is being squeezed as a consequence of the negative
square mass term in the equation of motion: (i.e crosses the Hubble radius): 
$l_{H}<\lambda$.
\end{itemize} 

A special feature of our dispersion relation is that, in Region I (RI), 
the non-linearity of the dispersion relation leads to a violation of the 
adiabaticity (WKB) condition.
This can be quantified in the following manner. We need to compare the 
rate of change of the physical frequency $\omega_{phys}$, 
namely $\frac{d}{d\eta}(\ln w_{phys})$, with the Hubble expansion rate 
$\CH \equiv a^{\prime}/a$. If the former is larger than the latter, then
adiabaticity is violated.
Thus, one defines an \emph{adiabaticity coefficient} \cite{Jerome3} by:
\beq
\label{E:alpha1}
\alpha(n,\eta) \, \equiv \, 
\left|\frac{\CH}{\frac{1}{w_{phys}}\frac{dw_{phys}}{d\eta}}\right| \, .
\eeq
If $\alpha \ll 1$ then the adiabaticity condition
is violated. The solution to the modified mode equation has no time to adjust itself 
such that an initial lowest energy state always tracks the instantaneous
lowest energy state of the time-dependent Hamiltonian of the mode.
Thus, even if the initial mode wave function minimizes the energy density,
it is possible that it is in an excited state when it enters the second region (RII). 
In this case, the initial conditions for the evolution of the mode in
Regions II and III will be different than what is assumed in the usual
linear theory of gravitational waves. Since short wavelength modes
spend more time in the region of non-adiabatic evolution than longer wavelength
ones, one could expect a change in the slope of the power spectrum of
cosmological fluctuations, as initially argued in \cite{Jerome1}.

For both power law inflation when 
\beq
a(\eta) \, = \, \frac{l_{0}}{\et^{1+\alpha}}
\eeq
with some constant $\alpha\geq 0$, and for exponential inflation when 
\beq \label{scalefactor}
a(\eta) \, = \, \frac{l_{0}}{\et} 
\eeq
we see that 
\beq
\CH \, \equiv \, \frac{a'}{a} \, = \, (1 + \alpha)\frac{1}{\et} \, .
\eeq
From Eqs. (\ref{E:4}) and (\ref{freq}), and neglecting the mass term for now, it
follows immediately that  
\beq 
\left|\frac{w_{phys}^{\prime}}{w_{phys}}\right| \, = \, \CH k l_{s} \, .
\eeq
Thus, the adiabaticity coefficient is
\beq 
   \alpha(n,\eta) \, \simeq \, \frac{\lambda}{l_S} \, ,
\eeq
where $\lambda$ is the physical wavelength corresponding to $k$. It
follows that the adiabaticity condition is violated in Region I and we should
hence expect modifications to the power spectrum.

\subsection{Minimum Energy Density Initial Conditions}

In Ref.\cite{Jerome1}, following Ref.\cite{Brown78}, it was argued 
that since the WKB approximation is not valid in the Trans-Planckian regime, 
we cannot choose as our initial conditions the Bunch-Davies vacuum. 
Instead we have to choose the state which initially minimizes the energy 
density of the field. With this prescription,
the initial value of the mode function becomes \cite{Jerome1} 
\beq \label{E:9}
\mu(\eta=\eta_{i}) \, = \, 
\frac{1}{\sqrt{2\omega(\eta_{i})}}=\frac{1}{\sqrt{2n_{eff}}} \, ,
\eeq
and from the minimal energy density requirement we obtain:
\beq
 \label{E:10}
 \mu^{\prime}(\eta=\eta_{i}) \, = \, \pm i \sqrt{\frac{n_{eff}}{2}} \, .
\eeq
Here, $\eta_{i}$ is the initial time at which we choose to minimize the 
energy density.

Other relevant mode-dependent times are:
\begin{itemize}
\item[(1)] $\eta_{1}$, the time when the mode labeled by a fixed comoving 
wavenumber $n$ crosses from Region I to Region II, is given by
\beq
\label{E:11}
|\eta_1|^{1 + \alpha} \, = \, {{l_0} \over {l_s}} {{2 \pi} \over n} \, 
\equiv \, \frac{2\pi}{n\epsilon} \, ,
\eeq
thus defining a constant $\epsilon \ll 1$.
\item[(2)] $\eta_{2}$, the time when the mode crosses the Hubble radius, i.e.
exits from Region II into Region III, is given by
 \beq
 \label{E:13}
 \etd \, = \, \frac{2\pi}{n}( 1 + \alpha)
 \eeq
\end{itemize}

\section{Spectrum in the Case of Minimal Energy Initial Conditions}

In this section, we present approximate analytical solutions for the mode
functions obtained using the modified dispersion relation derived above, and
compute the resulting power spectrum. For each mode, we solve the mode 
equation in each of the three regions (see Figure 1) and apply the standard
matching conditions at the boundaries between the regions.

\subsection{Region I (RI)}

Inserting the dispersion relation (\ref{E:8}) into the mode equation (\ref{EOM})
and making use of (\ref{E:4}) and of the form of the scale
factor for an exponentially expanding background (\ref{scalefactor})
- for simplicity we will focus on this case, i.e. on $\alpha = 0$ -
we obtain
%
\beq
\label{EOM1}
\mu^{\prime\prime}(\eta)+
\left[\frac{\exp(2n\epsilon|\eta|)}{\epsilon^{2}|\eta|^2}-\frac{a^{ \prime\prime}}{a}\right]\mu(\eta) \, = \, 0 \, .
\eeq

In Region I, the argument in the exponential in (\ref{EOM1})  is large, i.e.
$2 n \epsilon |\eta| > 1$, and hence we can approximate the equation of motion
in the following way: we can drop the second term - the squeezing term -
in the square parentheses of (\ref{EOM1}) since it is suppressed compared to the 
first term. In the first term, we replace $\eta$ by $\eta_1$ in the denominator
since the time dependence in the exponential overwhelms the time dependence
of the other factors. Thus, the approximate form of the
equation of motion Eq.(\ref{EOMR1}) in Region I becomes:
\beq
\label{EOMR1}
\mu^{\prime\prime}(\eta)\, + \, \frac{e^{2n\ep\et}n^2}{4\pi^2}\mu(\eta) \, = \, 0 \, .
\eeq
After the change of variables
\beq \label{xvalue}
x \, \equiv \, \frac{\exp(n\ep\et)}{2\pi\ep} \, ,
\eeq
the equation (\ref{EOMR1}) takes the form
%
\beq
\label{E:20}
x^2\mu^{\prime\prime} \, + \, x\mu^{\prime} \, + \, x^2\mu \, = \, 0 \, ,
\eeq
which is nothing but Bessel's differential equation of zeroth order.
The solution can be written as 
\beq
\mu(x) \, = \, AJ_{0}(x) \, + \, BY_{0}(x) \, ,
\eeq
where $J_0$ and $Y_0$ denote the Bessel functions and $A$ and $B$ are constants.

Since we are interested in the high energy behavior of this solution 
(recall that we are in the Trans-Planckian regime), we need the asymptotic 
representations for large arguments of the zeroth-order Bessel functions.
%
\be
\label{E:J}
J_{0}(x) \, \sim \, \sqrt{\frac{2}{\pi x}}\left[\cos(x-\frac{\pi}{4})\right]\,, & \mbox{$\forall \ x\gg 1$}\, \\
\label{E:Y}
Y_{0}(x) \, \sim \, \sqrt{\frac{2}{\pi x}}\left[\sin(x-\frac{\pi}{4})\right]\,, & \mbox{$\forall \ x\gg 1$}\,.
\en
Recalling the expression for $x$, the asymptotic solution $\mu = \mu_I$ in (RI) 
reads
\begin{widetext}
\be
\mu_{I}(\eta)=2\sqrt{\ep}\ e^{-n\ep\et/2}\left[A_1\cos\left(\frac{e^{n\ep\et}}{2\pi\ep}-\pi/4\right)+A_2\sin\left(\frac{e^{n\ep\et}}{2\pi\ep}-\pi/4\right)\right]
\en
\end{widetext}
Note that in terms of the variable $x$, the two distinguished mode-dependent
times $\eta_1$ and $\eta_2$ take the form
\be 
\label{E:X1}
x_1 \, &=& \, \frac{e^{n\ep\etu}}{2\pi\ep}=\frac{e^{2\pi}}{2\pi\ep}\gg1\\
\label{E:X2}
x_2 \, &=& \, \frac{e^{n\ep\etd}}{2\pi\ep}=\frac{e^{2\pi\ep}}{2\pi\ep}
\en
In terms of $x$, the solution reads
\beq
\label{E:mu(x)}
\mu_{I}(x) \, = \, 
\sqrt{\frac{2}{\pi x}}\left(A_1\cos(x-\pi/4)+A_2\sin(x-\pi/4)\right)
\eeq
The coefficients $A_1$ and $A_2$ are determined by the initial conditions.
We choose minimal energy density initial conditions. To implement these, we
compute the derivative of $\mu$ and use
\beq
\frac{d}{d\eta}\mu(\eta) \, = \, 
\frac{dx}{d\eta}\frac{d}{dx}\mu(x)=n\ep x\frac{d}{dx}\mu(x)
\eeq
From Eqs. (\ref{E:J}) and (\ref{E:Y})
\begin{widetext}
\beq
J_{0}^{\prime}(x) \, \simeq \, 
-n\ep\sqrt{\frac{2x}{\pi}}\left[\sin(x-\pi/4)+\frac{1}{2x}\cos(x-\pi/4)\right]
\eeq
and 
\beq
Y_{0}^{\prime}(x) \, \simeq \, 
n\ep\sqrt{\frac{2x}{\pi}}\left[\cos(x-\pi/4)-\frac{1}{2x}\sin(x-\pi/4)\right]
\eeq
and the asymptotic form of $\mu_I^{\prime}$ becomes
\beq
\label{E:mu'x}
 \mu_{I}^{\prime}(x) \, \simeq \, 
n\ep\sqrt{\frac{2x}{\pi}}\left[-A_1\sin(x-\pi/4)+A_2\cos(x-\pi/4)\right]
\eeq

To solve for $A_1$ and $A_2$ we need to apply the minimum energy density 
conditions Eqs. (\ref{E:9}) and (\ref{E:10}), but cast in terms of $x_{i}$ 
instead of $\eti$.
\be
\label{E:matchI}
A_1\cos(x_i-\pi/4)+A_2\sin(x_i-\pi/4) \, &=& \, \sqrt{\frac{\pi x_i}{4n_{eff_{i}}}}\\
\label{E:matchII}
-A_1\sin(x_i-\pi/4)+A_2\cos(x_i-\pi/4) \, &=& \, \pm\frac{i}{n\ep}\sqrt{\frac{\pi n_{eff_i}}{4x_i}}
\en 
We can now use the fact that $\ep\ll1$ and the condition that $n\ep\et\gg1$ 
to guide us through our approximations.
Solving for $A_1$ and $A_2$ we get
\be
A_1 \, &=& \, \sqrt{\frac{\pi x_i}{4n_{eff_{i}}}}\cos(x_i-\pi/4)\mp\frac{i}{n\ep}\sqrt{\frac{\pi n_{eff_i}}{4x_i}}\sin(x_i-\pi/4)\\
A_2 \, &=& \, \sqrt{\frac{\pi x_i}{4n_{eff_{i}}}}\sin(x_i-\pi/4)\pm\frac{i}{n\ep}\sqrt{\frac{\pi n_{eff_i}}{4x_i}}\cos(x_i-\pi/4)
\en
Comparing the two terms in each expression, we find that
\be
\label{E:A1}
A_1 \, &\simeq& \, \sqrt{\frac{\pi x_i}{4n_{eff_{i}}}}\cos(x_i-\pi/4)\\
\label{E:A2}
A_2 \, &\simeq& \, \sqrt{\frac{\pi x_i}{4n_{eff_{i}}}}\sin(x_i-\pi/4)
\en
Eq. (\ref{E:mu(x)}) thus gives us the following solution in Region I
\beq
\mu_{I}(x) \, = \, \frac{1}{\sqrt{2n_{eff_i}}}\sqrt{\frac{x_i}{x}}
\left[\cos(x_i-\pi/4)\cos(x-\pi/4)+\sin(x_i-\pi/4)\sin(x-\pi/4)\right]
\eeq
\end{widetext}
or, more compactly,
\beq
\mu_{I}(x) \, = \, \frac{1}{\sqrt{2n_{eff_i}}}\sqrt{\frac{x_i}{x}}\cos(x_i-x)
\eeq

\subsection{\bf{Region II (RII)}}

In Region II, the adiabaticity condition is not violated, and thus the
mode functions $\mu = \mu_{II}$ are oscillatory, i.e. 
\beq
\label{E:39}
\mu_{II}(\eta) \, = \, B_{1}e^{in\eta}+B_{2}e^{-in\eta}
\eeq
or equivalently in terms of $x$
\be 
\label{E:muII(x)}
\mu_{II}(x) \, = \, 
B_1(2\pi\ep)^{i/\ep}x^{i/\ep}+B_2(2\pi\ep)^{-i/\ep}x^{-i/\ep}\\
\label{E:muII'(x)}
\mu_{II}^{\prime}(x) \, = \, 
in\left[B_1(2\pi\ep)^{i/\ep}x^{i/\ep}-B_2(2\pi\ep)^{-i/\ep}x^{-i/\ep}\right] \, .
\en

The coefficients $B_{1}$ and $B_{2}$ are determined by the matching 
conditions at the time $\etu=\frac{2\pi}{n\ep}$
\be
\label{E:40}
\mu_{I}(\etu)=\mu_{II}(\etu)\nonumber\\
\mu^{\prime}_{I}(\etu)=\mu^{\prime}_{II}(\etu)
\en
or their equivalent in terms of the variable $x$. Recall that, in terms of
the $x$ variable,
$\mu_{I}(x_1)$ and $\mu_{I}^{\prime}(x_1)$ are given by
\be
\mu_{I}(x_1) \, &=& \, \sqrt{\frac{\ln|2\pi\ep x_i|}{4\pi n\ep}}x^{-1/2}\cos(x_1-x_i)\\
\mu_{I}^{\prime}(x) \, &\simeq& \, \sqrt{\frac{n\ep\ln|2\pi\ep x_i|}{4\pi}}\sin(x_i-x_1) \, ,
\en 
where we have again used the fact that $\ep\ll1$ and $n\ep\et\gg1$. 
To abbreviate the notation, we introduce the variable
\beq \label{calA}
\CA_i \, \equiv \, \sqrt{\ln|2\pi\ep x_i|} \, .
\eeq
Then, the matching conditions (\ref{E:40}) (making use of the 
explicit value of $x_1$ from Eq.(\ref{E:X1})) read
\begin{widetext}
\be
\frac{e^{-\pi}}{\sqrt{2n}}\CA_i\cos(x_i-x_1) \, = \, 
B_1(2\pi\ep)^{i/\ep}x_{1}^{i/\ep}+B_2(2\pi\ep)^{-i/\ep}x_{1}^{-i/\ep}\\
\frac{-ie^{\pi}}{\sqrt{2n}}\frac{\CA_i}{2\pi}\sin(x_i-x_1) \, = \, 
B_1(2\pi\ep)^{i/\ep}x^{i/\ep}-B_2(2\pi\ep)^{-i/\ep}x^{-i/\ep}
\en
The coefficients $B_1$ and $B_2$ are thus given by
\be
B_1 \, = \, 
\frac{\CA_i}{2\sqrt{2n}}((2\pi\ep)^{-i/\ep}x_{1}^{-i/\ep}\left[e^{-\pi}\cos(x_i-x_1)-i\frac{e^{\pi}}{2\pi}\sin(x_i-x_1)\right]\\
B_2 \, = \, 
\frac{\CA_i}{2\sqrt{2n}}((2\pi\ep)^{i/\ep}x_{1}^{i/\ep}\left[e^{-\pi}\cos(x_i-x_1)+i\frac{e^{\pi}}{2\pi}\sin(x_i-x_1)\right] \, .
\en
This time, we must keep all the terms in the coefficients since they are all 
of the same order. Thus,
the solution of the mode function in (RII), Eq.(\ref{E:muII(x)}), reads
\beq
\mu_{II}(x) \, = \, 
\frac{\CA_i}{2\sqrt{2n}}\left[\left(\frac{x}{x_1}\right)^{i/\ep}\left(e^{-\pi}\cos(x_i-x_1)-i\frac{e^{\pi}}{2\pi}\sin(x_i-x_1)\right)+\left(\frac{x}{x_1}\right)^{-i/\ep}\left(e^{-\pi}\cos(x_i-x_1)+i\frac{e^{\pi}}{2\pi}\sin(x_i-x_1)\right)\right]\nonumber
\eeq
or, more compactly,
\beq
\label{E:muII}
\mu_{II}(x) \, = \, 
\frac{\CA_i}{\sqrt{2n}}\left[e^{-\pi}\cos(x_i-x_1)\cos\left(\frac{\ln|x/x_1|}{\ep}\right)+\frac{e^{\pi}}{2\pi}\sin(x_i-x_1)\sin\left(\frac{\ln|x/x_1|}{\ep}\right)\right]
\eeq
\end{widetext}

\subsection{\bf{The Power Spectrum}}

Now we are in a position to compute the power spectrum. Assuming ${\cal O}(1)$
coupling between the growing mode in Region III and $\mu_{II}$, the 
non-decaying mode mode in Region III is
\beq
\label{E:48}
\mu_{III}(\eta) \, \simeq \, Ca(\eta) 
\eeq
where the constant $C$ is fixed by matching $\mu_{III}$ with $\mu_{II}$
at the Hubble-crossing time $\etd$ (see Eq.(\ref{E:13}))
\beq
\mu_{III}(\etd) \, = \, Ca(\etd) \, = \, C\frac{l_0}{\etd}
\eeq
In terms of the variable $x$ the matching conditions are
\beq
\mu_{III}(x_2) \, = \, Cn\frac{l_0}{2\pi} \, = \, \mu_{II}(x_2) \, ,
\eeq
and hence
\beq \label{Cvalue}
C \, = \, \frac{2\pi}{nl_0}\mu_{II}(x_2) \, .
\eeq
						
The power spectrum $P_{\phi}$ for our scalar field $\phi$ is defined as
\beq \label{powerspec}
P_{\phi}(n) \, = \, n^3 |\phi(k)|^2 \, = \, n^3 a^{-2} |\mu(n)|^2 \, .
\eeq
The spectral index $n_T$ for gravitational waves is defined by 
\beq
\label{specindex}
P_{\phi}(n) \, = \, A_{\rm T}n^{n_{\rm T}} \, ,
\eeq
where $A_{\rm T}$ is the amplitude of the spectrum.

Combining (\ref{powerspec}) and (\ref{E:48}) we obtain
\beq
P_{\phi}(n) \, = \, n^{3}\left|\frac{\mu}{a}\right|^{2} \, = \, n^{3}|C|^{2} \, .
\eeq
Making use of (\ref{Cvalue}) and (\ref{E:muII}) we get
\be
P_{\phi}(n) \, &=& \, \frac{(2\pi)^2 n}{l_{0}^{2}}\left|\mu_{II}(\etd)\right|^{2}
\\
&=& \, \frac{2\pi^2}{l_{0}^{2}} |\CA_i|^2 \CJ(\eti,\etu,\etd)|^2
\en
\begin{widetext}
where $\CJ(\eti,\etu,\etd)$ will yield oscillations in the spectrum of a
particular form, and is given by
\be
\label{E:CJ}
\CJ(\eti,\etu,\etd) \, = \, 
e^{-\pi}\cos\left(\frac{e^{n\ep\eti}-e^{n\ep\etu}}{2\pi\ep}\right)\cos\left(\frac{n\ep\etd}{n\ep\etu}\right)
+\frac{e^{\pi}}{2\pi}\sin\left(\frac{e^{n\ep\eti}-e^{n\ep\etu}}{2\pi\ep}\right)\sin\left(\frac{n\ep\etd}{n\ep\etu}\right)
\en

Plugging in the values for $\etu$ and $\etd$ and making more transparent the 
$n\ep\eti$-dependence of $\CJ(n,\ep,\eti)$, we obtain 
\be
\label{E:CJI}
\CJ(n,\ep,\eti) \, = \, \cos\left(\frac{e^{n\ep\eti}}{2\pi\ep}\right)\left[e^{-\pi}\cos\left(\frac{2\pi}{\ep}\right)\cos\left(\frac{e^{2\pi}}{2\pi\ep}\right)-\frac{e^{\pi}}{2\pi}\sin\left(\frac{e^{2\pi}}{2\pi\ep}\right)\right] \nonumber\\
+ \sin\left(\frac{e^{n\ep\eti}}{2\pi\ep}\right)\left[e^{-\pi}\cos\left(\frac{2\pi}{\ep}\right)\sin\left(\frac{e^{2\pi}}{2\pi\ep}\right)-\frac{e^{\pi}}{2\pi}\cos\left(\frac{e^{2\pi}}{2\pi\ep}\right)\right]
\en
\end{widetext}

The amplitude and spectral index of the power spectrum can be read off
by inserting the expression (\ref{calA}) for $\CA_i$ and making use of
the definition (\ref{xvalue}) of the variable $x$, yielding
\beq
\label{E:PowerI}
A_{\rm T}n^{n_{\rm T}} \, = \, 
\frac{2\pi^{2}}{l_{0}^{2}}\ n\ep\eti \ |\CJ(n,\ep,\eti)|^2 \, .
\eeq

\subsection{\bf{Discussion}}\label{S:DiscI}

From our main result (\ref{E:PowerI}) we draw the following
conclusions. Given our modified dispersion relation for fluctuations,
the spectrum of gravitational waves in an exponentially expanding
background is characterized by a non-vanishing spectral tilt of
$n_{\rm T} = 1$. In addition, there are oscillations in the
spectrum with a characteristic dependence on $n$. This result was
obtained working to leading order in $\ep$.

The fact that the spectrum is not scale-invariant but blue-shifted  
is due to the fact that the shorter wavelengths are subject to the
modified dispersion relation for a longer time, leading to a higher
excitation level of the modes when their wavelength is equal to the
cutoff scale after which the dispersion relation becomes linear.
This is similar to what happens in models with a discrete space-time, or
in analog models from condensed matter physics where the modes feel the 
granular nature of matter on scales of the order of the atomic separation 
resulting  in a departure from the linear dispersion relation. 

We would like to point to an important observation regarding Eq.(\ref{E:PowerI}). 
We made it explicit to show how the combination $n\ep\eti$ appears both in 
the oscillatory function $\CJ$ and in Eq.(\ref{E:PowerI}). We see that to 
get a scale invariant spectrum without oscillations, it suffices to 
modify the prescription for the initial conditions and postulate that
the modes are created at a mode-dependent initial time $\eti$ for which 
$n\eti\equiv$\emph{constant} (this is similar to what is postulated to
occur in the analyses of \cite{minmal,Ho}).
We will discuss later on a better derivation of what the initial conformal 
time might be by linking the fact that having T-duality at the basis of our 
dispersion relation, i.e. as our Trans-Planckian physics, is similar as 
modifying our space-time metric by adding to it a minimal length which is 
of the order of the string length $l_S$. This could be linked to 
\emph{non-commutative space-time} or to a 
\emph{stringy space-time uncertainty relation} which was studied earlier 
in Ref.\cite{Ho}. We will return to that point at the 
end of our article since it is of a more speculative nature.

\section{Spectrum in the Case of an Instantaneous Minkowski Vacuum}

To test the dependence of our previous result on the \emph{initial conditions} 
we adopt another choice of initial conditions, the 
\emph{instantaneous Minkowski vacuum} as an alternative to Eqs.(\ref{E:9}) 
and (\ref{E:10}). That is, at $\et=\eti$ the mode function now satisfy:
\be
\label{E:MinK1}
\mu(\eti)=\frac{1}{\sqrt{2n}}\\
\mu'(\eti)=\pm\sqrt{\frac{n}{2}}
\en
\subsection{\bf{The Trans-Planckian Region (RI)}}

Following the same steps as in the case of the 
\emph{minimum energy density initial state}, we set the initial conditions 
at a fixed time $\eta_i$ in the Trans-Planckian region (RI). We start
with the general solution in Region I from Eqs.(\ref{E:mu(x)}) 
and (\ref{E:mu'x}) and match them to the initial values (\ref{E:MinK1})
in order to determine the new coefficients $A_1$ and $A_2$.
The result is
\be
\label{E:A1Mink}
A_1 \, \simeq \, \sqrt{\frac{\pi x_i}{4n}}\cos(x_i-\pi/4)\\
\label{E:A2Mink}
A_2 \, \simeq \, \sqrt{\frac{\pi x_i}{4n}}\sin(x_i-\pi/4) \, ,
\en
and the mode function $\mu_{I}(x)$ is thus given by
\be
\mu_{I}(x) \, = \, \frac{1}{\sqrt{2n}}\sqrt{\frac{x_i}{x}}\cos(x_i-x) \, .
\en

\subsection{\bf{Region II}}

In Region II where the WKB approximation is valid, we have plane
wave solutions for the mode functions as in Eqs. (\ref{E:muII(x)}) 
and (\ref{E:muII'(x)}). 
The mode functions are matched at the time $\etu$ when the physical 
wavelength of a mode equals the minimal length $l_S$, 
that is
\be
\label{E:MI(x1)}
\mu_{I}(x_1) \, = \, \mu_{II}(x_1)\\
\label{E:MI'(x1)}
\mu_{I}'(x_1) \, = \, \mu_{II}'(x_1)
\en
These yield two equations for the coefficients $B_1$ and $B_2$ which
can be solved to get
\be
B_1 \, \simeq \, 
\frac{-i}{2}\sqrt{\frac{x_{i}x_{1}}{2n}}\frac{\sin(x_i-x_1)}{(2\pi\ep x_1)^{i/\ep}}\\
B_2 \, \simeq \, 
\frac{i}{2}\sqrt{\frac{x_{i}x_{1}}{2n}}\sin(x_i-x_1)(2\pi\ep x_1)^{i/\ep}
\en
The mode solution $\mu_{II}(x)$ in Region II is thus given by:
\beq
\label{E:muII(x)Mink}
\mu_{II}(x) \, = \, 
\sqrt{\frac{x_{i}x_{1}}{2n}}\sin(x_i-x_1)\sin\left[\frac{\ln|x/x_1|}{\ep}\right]
\, .
\eeq

\subsection{\bf{The Power Spectrum Revisited}}

Again assuming that the growing mode in Region III picks up ${\cal O}(1)$
of the amplitude of $\mu_{II}$ at the matching time $x_2$, the solution
$\mu_{III}$ in Region III is given by
\beq
\mu_{III} \, = \, C a(\eta) \, ,
\eeq
with the constant $C$ determined by
$C=\frac{2\pi}{l_{0}n}\mu_{II}(x_2)$ or, expanded out
\beq
C \, = \, \frac{2\pi}{l_{0}n}\sqrt{\frac{x_{i}x_{1}}{2n}}\sin(x_i-x_1)\sin\left[\frac{\ln|x_2/x_1|}{\ep}\right]
\eeq
The power spectrum $P_{\phi }(n)$ thus becomes
\be
P_{\phi}(n) \, &=& \, A_Sn^{n_T} \, = \, n^{3}|C|^{2} \\
&=& \, \frac{2\pi^{2}}{l_{0}^{2}}x_{i}x_{1}\sin^{2}(x_i-x1)\sin^{2}\left[\frac{\ln|x_2/x_1|}{\ep}\right]
\en
In terms of conformal time, 
\beq
\label{E:PowerII}
P_{\phi}(n) \, = \, \frac{e^{2\pi}e^{n\ep\eti}}{2l_{S}^{2}}\sin^{2}\left[\frac{e^{n\ep\eti}-e^{2\pi}}{2\pi\ep}\right]\cos^{2}\left(\frac{2\pi}{\ep}\right)
\eeq

\subsection{\bf{Discussion}}

The striking difference between our previous result obtained using 
the minimum energy density calculations and that found here using the 
instantaneous Minkowski vacuum initial conditions is the dependence of the 
power spectrum on the comoving wave number $n$. We see from Eq. (\ref{E:PowerII}) 
that the dependence on $n$ is exponential while from Eq. (\ref{E:PowerI}) the 
dependence on $n$ is linear. Thus, for the instantaneous Minkowski vacuum
we get an exponentially blue spectrum 
(note that the quantity $n\ep\eti$ which appears in the exponential
is much greater than one). 

If, instead of imposing initial conditions for all modes at a fixed
time, we use the alternative discussed section \ref{S:DiscI}, namely 
assume a minimal length in our theory and argue that initial 
conditions should be set on a surface of constant $n\eti$ which would 
correspond to the surface on which the physical mode are created,
and setting  
\beq
n\eti \, = \, \frac{\ln|2\pi^{2}\ep\delta|}{\ep} \, ,
\eeq
(for some fixed number $\delta$) we would obtain the following spectrum:
\beq
 A_{\rm S}n^{n_{\rm S}-1} \, = \, 
\frac{e^{2\pi}\pi^{2}\delta}{l_{S}^{2}}\sin^{2}\left(\pi\delta-\frac{e^{2\pi}}{2\pi\ep}\right)\cos^{2}\left(\frac{2\pi}{\ep}\right)
\eeq
which is scale invariant spectrum without any oscillations which depend on the 
comoving wavenumber $n$. Thus, for this initial time prescription we arrive 
at the same conclusion as in the case of minimal energy initial conditions. 
This is not a real surprise since in this case all modes spend the same
amount of time in the trans-Planckian region.

\section{Conclusions}

In this paper we have studied the spectrum of gravitational waves which
results when the usual linear dispersion relation for the wave equation
is replaced by the non-linear dispersion relation which results when
assuming that the wave propagator is consistent with the T-duality
symmetry of string theory. This modified dispersion relation differs
from the linear one on length scales smaller than the string scale.

We have shown that the modified dispersion relation leads to a non-adiabatic
evolution of the mode functions in the trans-Planckian region. If, as is
usual in an inflationary cosmology, the initial conditions for fluctuations
are set for all modes at the same initial time, this modification, in
turn, leads to a change in the slope of the spectrum of gravitational
waves. Instead of a scale-invariant spectrum, a spectrum with blue tilt
given by a spectral index $n_T = 1$ results. In addition, the spectrum
has characteristic oscillations.

Our results are based on studying the evolution of a test scalar field
on our expanding inflationary background. According to the theory of
cosmological perturbations, each gravitational wave polarization mode
obeys the same equation as such a test scalar field. The scalar metric
fluctuations, in turn, obey an equation with a slightly different
squeezing factor \cite{MFB,RHBrev2}. 
The difference in the equations of motion, however,
is in general only important on scales larger than the Hubble radius,
and even then only if the equation of state of the background is
changing in time. Thus, it is reasonable to expect that a similar analysis
to the one given in this paper applies to scalar metric fluctuations.
The result for scalar metric fluctuations would then be a spectrum with
blue tilt $n_S = 2$, inconsistent with recent data \cite{WMAP}. If the
only change to the equation of motion for cosmological fluctuations
in the context of string theory on trans-Planckian scales were the
change in the dispersion relation discussed in this paper, our results
would imply a deep problem in realizing successful inflation in the
context of string theory. However, as discussed in \cite{Ho},
there are other aspects of string theory, e.g. the space-time
uncertainty relation, which must be considered, and these considerations
could well restore the prediction of a scale-invariant spectrum in the
case when the background space in exponentially expanding.

In conclusion, we hope to have convinced the reader that, assuming that
there was a period of cosmological inflation, the basic principles
of string theory are clearly testable in cosmological observations.
Final predictions, however, will have to await a better understanding
of the equations of string theory on trans-Planckian scales.

\begin{acknowledgments}

This research has been supported by an NSERC Discovery Grant, and by
the Canada Research Chairs program.

\end{acknowledgments}

\end{document}